\documentclass[%
 reprint,
superscriptaddress,
groupedaddress,
preprintnumbers,
nofootinbib,
 amsmath,amssymb,
 aps,
dvipsnames
]{revtex4-2}

\usepackage{graphicx}
\usepackage{dcolumn}
\usepackage{bm}
\usepackage[normalem]{ulem} 
\usepackage{tikz}
\usetikzlibrary{decorations.pathmorphing}
\usetikzlibrary{decorations.markings}
\usetikzlibrary{calc}
\usepackage{verbatim}
\usepackage{caption}
\usepackage{subcaption}

\newcommand{\Amp}[1]{{\mathcal A}_{ #1}}
\newcommand{\vof}[1]{{\langle{ #1}\rangle}}
\newcommand{\bh}{{\bf h}}

\begin{document}
\preprint{IPPP/21/29}
\title{On the road(s) to the Standard Model}
\author{Rodrigo Alonso} \author{Mia West} 
\affiliation{Institute for Particle Physics Phenomenology, Durham University, South Road, Durham, DH1 3LE}

\begin{abstract}
Experimental measurements point at the Standard Model (SM) as the theory of electroweak symmetry breaking, but as we close in on our characterization the question arises of what limits in theory space lead to the SM. The class of theories with this property cannot be ruled out, only constrained to an ever smaller neighbourhood of the SM. In contrast, which class of theories do not posses this limit and can therefore potentially be ruled out experimentally? In this work we study both classes and find evidence to support the Standard Model Effective field Theory being the single road to the Standard Model, theories that fall outside this class keeping a `minimum distance' from the SM characterized by a cut-off of at most $ 4\pi v/g_{\rm SM}$.
\end{abstract}

\maketitle

\section{Introduction}

Nowadays, particle physics finds itself in the midst of Electroweak Symmetry Breaking (EWSB) exploration, the outcome of this endeavour will chart Nature's theory of elementary particles. Experimental data collated and compared with predictions of theories of EWSB has narrowed down the range possibilities; many a casualties lie indeed now discarded having been disproven by the progress in our measurements. The Higgs boson discovery, coming up on a decade old, was the main stroke in our map, subsequent data giving a profile that resembles the one heralded by the Standard Model (SM). Theory considerations have long pointed out the SM case for EWSB to be unstable under higher scale corrections and indicated that new physics should lie in wait at the electroweak scale. Whether these considerations should be revisited and our theory perspective profoundly changed, or if instead patience is all that is needed, the pressing question at present posed by experimental data is to characterize the theory `neighbourhood' of the SM. The claim that one observes nothing but the SM at the LHC is indeed only as good our characterization of what else we could observe; it is here we find value in the aforementioned casualties. The aim in this work is to explore the consistent theory neighbourhood of the Standard Model. 

A long known and studied approach, or `trajectory', to the SM is a linearly realized Effective Field Theory (SMEFT), see \cite{Brivio:2017vri} for a review, this road being pointed at by the decoupling theorem \cite{Appelquist:1974tg}. The integration of any heavy particle whose mass can be arbitrarily larger than the EWSB vev ($M>v$) in a perturbative linear realization will yield the SMEFT; supersymmetry or composite Higgs models fall into this category.
Is this the only road to the Standard Model, i.e. are there other consistent limits to obtain the SM couplings for the known spectrum of elementary particles? As fundamental as this topic is, on its present formulation the candidate preceded the question; 
Higgs Effective Theory~\cite{Feruglio:1992wf,Grinstein:2007iv} is an EFT that encompasses the SMEFT but extends beyond it and might offer new roads. In HEFT, a linear realization is not assumed (though admissible in certain limit) and is indeed the most general Lorentz and gauge invariant theory with the known spectrum of particles (which suggests it should be possible to formulate it in terms of amplitudes). The theories that this EFT describes but fall out of SMEFT, which will be called here theories of the quotient space HEFT/SMEFT or simply quotient EFTs~\footnote{In other works these are called, with a slight abuse of notation, HEFT.}, could contain a path to the SM other than via SMEFT. This quotient space is characterized as missing a point in field space which is left invariant under an $O(4)$ transformation~\cite{Alonso:2015fsp,Alonso:2016oah}, be it because it is not present or because the would be invariant point is singular~\cite{Cohen:2020xca}. A geometric formalism was used to derive this result and also aids in exploring the properties of theories without field redundancies, as introduced in \cite{Alonso:2015fsp,Alonso:2016oah}, and followed up in \cite{Cohen:2020xca,Helset:2020yio,Cohen:2021ucp} - it is also adopted here.
 Some theories in HEFT/SMEFT quotient space have been formulated while having a perturbative expansion~\cite{Cohen:2020xca}; they have been found to have a cut-off of  $\sim 4 \pi v$ and no limit can be taken within them that yields the SM. It has been suggested that {\it all} of this this quotient space shares this property of a finite $v$-bound cut-off~ \cite{Falkowski:2019tft} with further evidence provided in~\cite{Cohen:2021ucp}, which means in turn that they all could be casualties of our exploration with present and future machines. This question has been explored so far with perturbative unitarity bounds, while here it is looked at with semi-classical arguments.

This letter is structured as follows. Section \ref{SecII} introduces geometry from amplitudes, and sec.~\ref{SecIIA} presents the basis in Riemann normal coordinates. This first part has been rendered review, rather than new results, by virtue of~\cite{Cohen:2021ucp} although all results here are derived independently. 
Section.~\ref{SecIIB} presents theory and experimental bounds on the curvature plane while \ref{SecIII} characterizes SMEFT on this plane. In sec.~\ref{SecIV}, example models of SMEFT and quotient space are presented and characterized in the curvature plane. 
Sec.~\ref{SecV} presents theories in quotient space arising from geometry rather than explicit models and finds candidate quotient theories that seem to approach the SM. A semi-classical argument for the finite cut-off of theories in quotient space is given in \ref{SecVI}.


\section{Geometry and Amplitudes}\label{SecII}

For simplicity, $O(4)\supset SU(2)\times U(1)$ invariance in the EWSB sector is assumed. We take the high energy limit and make use of the equivalence theorem. These approximation allow us to focus on a subset of possible modifications of the bosonic sector. The Higgs singlet field is denoted $h$, and the Goldstones swallowed by the $W$ and $Z$ bosons as $\varphi^a$, $a=1,2,3$. 

Let us start by defining our geometry from the scattering-matrix $S$ in order to depart from a common-place, basis-invariant magnitude in particle physics. Following the line-integral definition for general amplitudes valid also in the UV, we have ($S=1-i \mathcal A$):
\begin{align}\label{Rphi-A}
-R_{h+h-}&=\frac{1}{2\pi i}\oint\frac{1}{s_{12}^2} \mathcal A_{W^+_1W^-_2\to hh}\\
	-R_{+-+-}&= \frac{1}{2\pi i}\oint \frac{1}{s_{12}^2}\mathcal A_{W^+_1W^+_2\to W^+W^+}\\
	-\nabla_h R_{+h-h}&=\frac{1}{2 \pi i}\oint \frac{1}{s_{12}^2}\mathcal A_{W_1^+W_2^-\to hhh}\\
	-\nabla_{h} R_{+-+-}&=\frac{1}{\pi i}\oint \frac{1}{s_{12}^2}\mathcal A_{W_1^+W_2^+\to W_3^+W_4^+h}\\
	&=\frac{1}{\pi i}\oint \frac{1}{s_{34}^2}\mathcal A_{W_1^+W_2^+\to W_3^+W_4^+h} \label{DR-Aww}
\end{align}
where $s_{ij}=(p_i+p_j)^2$. Indices in the Riemann tensor run through $h, a=1,2,3$ and the $\pm$ entries are given by contracting an $a$-index with the projector $(\delta^a_{\,1}\pm i\delta^{a}_{\,2})/\sqrt{2}$, for example
\begin{align}
	R_{h+h-}&=R_{hahb}\frac{(\delta^a_{\,1}+i\delta^a_{\,2})}{\sqrt{2}}\frac{(\delta^b_{\,1}-i\delta^b_{\,2})}{\sqrt{2}}
\end{align}
While the above definition is useful to include UV models and derive positivity bounds~\cite{Falkowski:2012vh}, in practice we will work with the low energy EFT. In which case the correspondence is taking our geometry from the order $\mathcal O(s)$ coefficients in a Taylor expansion. What's more is they capture all terms to this order. Being explicit,
\begin{align}\label{AWW}
	\mathcal A_{W_1^+W_2^-\to hh}=& - s_{12}R_{+h-h} \\ \label{AWh}
	\mathcal A_{W^+_1W^+_2\to WW}=&-s_{12}R_{+-+-} \\
	\mathcal A_{W_1^+W_2^-\to hhh}=&-s_{12}\nabla_h R_{+h-h} \\
	\mathcal A_{W^+_1W^+_2\to W_3^+W_4^+ h}=&-\frac{s_{12}+s_{34}}{2}\nabla_h R_{+-+-}
\end{align}
where we neglected masses assuming $s\gg M_W^2,M_Z^2,m_h^2$.

This starting point makes evident that our tensor, $R$, and its derivatives are physical and field redefinition (coordinate) invariant. Even if intuitive, this last statement should be qualified. On the geometry side, having defined tensor entries rather than invariants, one has that these change under coordinate transformations - albeit with well defined properties. They are nonetheless the same for local (defined around the vacuum) transformations of our fields which leave the amplitudes the same~\cite{PhysRev.177.2239}:
\begin{align}
	\hat \phi^i =& \left(\delta^i_j+\sum_{k=1} c^k_{j} \phi^k\right)\phi^j
\end{align}
so that after quantization both fields produce a particle out of the vacuum, 
\begin{align}
	\langle p| \phi^i| 0\rangle = \langle p| \hat \phi^i| 0\rangle
\end{align}
 with $|p\rangle$ the state associated with the field. It is for this type of transformation that the $S$ matrix will be left invariant, and tensors evaluated at the vacuum transform trivially, since:
 \begin{align}
 \left.\frac{\partial \phi^i}{\partial \hat\phi^j}\,\right|_{\phi=0}=\delta^i_j
 \end{align}
 
 Still, from where we stand the definition of Riemann tensor components in terms of amplitudes seems arbitrary and potentially inconsistent. So let us now turn to the Lagrangian theory which yields such relations.

\subsection{Riemann Normal Coordinates} \label{SecIIA}

Take the metric that the Riemann tensor derives from in eqs.~(\ref{Rphi-A}-\ref{DR-Aww}) as $G_{ij}(\phi)$, with $i,j=h,1,2,3$, $\phi=(h,\varphi^a)$ $a=1,2,3$. The amplitudes in eqs.~(\ref{Rphi-A}-\ref{DR-Aww}) follow from the action
\begin{align}\nonumber
S=&\frac 12\int d^4x \partial_\mu \phi^i G_{ij}\partial^\mu\phi^i\\
=&\frac 12\int d^4x\left( \partial_\mu h \partial^\mu h+F(h)^2g_{ab}\partial^\mu \varphi^a\partial_\mu \varphi^b\right) \label{OurL}
\end{align}
In matrix notation, our parametrization of the metric reads
\begin{align}
G_{ij}=\left(\begin{array}{cc}
1&\\
&F^2 g_{ab}
\end{array}\right)
\end{align}
where off-diagonal entries are forbidden by symmetry and $g_{ab}$ is the metric on the 3-sphere which we find useful to represent via the unit vector $u(\varphi)$: 
\begin{align}
	g_{ab}=& \frac{\partial u(\varphi)}{\partial \varphi^a}\frac{\partial u(\varphi)}{\partial \varphi^b} & u\cdot u&=1
\end{align}
with $u$ transforming as a vector under $O(4)$.
It follows that the non-vanishing elements of the Riemann tensor and its first covariant derivative are
\begin{align}
	\label{EqRabcd}
	R_{abcd}&=\left(\frac{1}{v^2}-(F')^2 \right)F^2 g_{a[c}g_{bd]}
	\\ \label{EqRahbh}
	R_{ahbh}&=-F''F \tilde g_{ab}
	\\
	\nabla_h R_{ahbh}&=F^2 \left(-\frac{F''}{F}\right)' g_{ab}\\
	\nabla_h R_{abcd}&=F^4 \left(\frac{1}{v^2F^2}-\frac{(F')^2}{F^2} \right)' g_{a[c}g_{bd]} \\
	\nabla_a R_{hbcd}&=\frac{F^4}{2} \left(\frac{1}{v^2F^2}-\frac{(F')^2}{F^2} \right)' g_{a[c}g_{bd]}
\end{align}
where prime denotes differentiation with respect to $h$ and it is useful to define
\begin{align}\label{Rdef}
	R_h&\equiv-\frac{F''}{F} & R_\varphi&\equiv\frac{1}{v^2F^2}-\frac{(F')^2}{F^2}
\end{align}

\begin{figure}
\includegraphics[width=\linewidth]{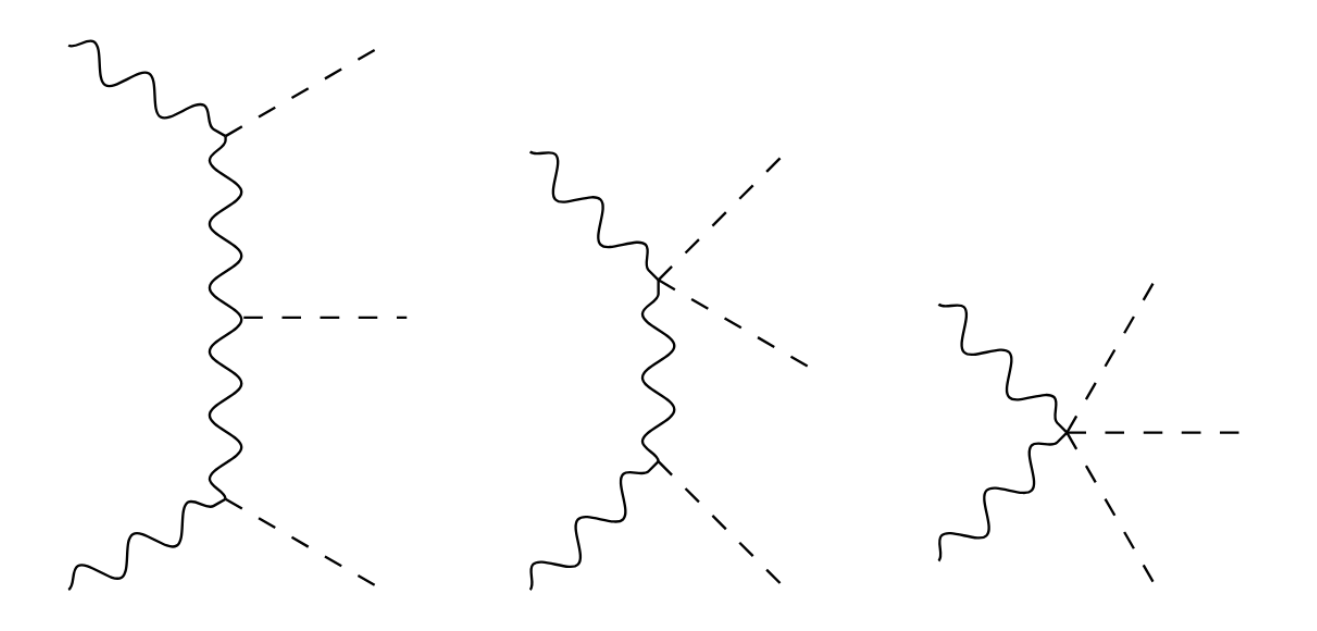}
\caption{\label{fig5pt} Diagrams for the $\mathcal O(s)$ contribution to the $WWhhh$ amplitude in the basis of eq.~(\ref{OurL}).}
\end{figure}

Verifying that these tensor entries appear as coefficients in the 4- and 5-point amplitudes is a matter of computing amplitudes: expanding our metric around the vacuum and adding over the various diagrams, e.g. see fig.~\ref{fig5pt} for those contributing to $WW\to hhh$, relations~(\ref{Rphi-A}-\ref{DR-Aww}) are recovered. The $O(4)$ symmetry in our system reduces the number of independent components and amplitudes to $R_h$, $R_\varphi$ and its derivatives.

Geometry does tell us however, that there is a frame where this computation is particularly simple: the frame where our coordinates follow geodesics, i.e. Riemann normal coordinates (RNC).
 
Let us then go into a brief outline of RNC. One can solve iteratively the Geodesic equation:
\begin{align}
	\frac{d^2 \phi^i}{d\sigma^2}+\Gamma^i_{jk}(\phi)\frac{d\phi^j}{d\sigma}\frac{d\phi^k}{d\sigma}=0
\end{align} 
 in an expansion which assumes the dependence on $\phi$ of $\Gamma$ admits a Taylor expansion and introduces new coordinates $\phi'$ defined to second order as
 $$
 \phi'^{i}=\phi^i+\frac12 \Gamma^{i}_{jk}(0) \phi^j\phi^k+\mathcal O(\phi^3)
 $$
 Together with a metric in the new coordinates and to $\phi'^3$ order~\cite{Hatzinikitas:2000xe}:
 $$
 G(\phi')_{ij}=G(0)_{ij}+\phi'^k\phi'^l\frac13 R_{iklj}+\frac16\phi'^{k}\phi'^l\phi'^m\nabla_m R_{iklj}
 $$
 For concreteness, one can work out this transformation for our metric to find:
 \begin{align}\nonumber
 	\left(\begin{array}{c}
 	 h'\\ \varphi'
 	\end{array}\right)=\left(\begin{array}{c}
 	 h-FF' \varphi^2/2\\  \varphi^a+ F'h\varphi^a/F+\Gamma^a_{bc}\varphi^b\varphi^c/2
 	\end{array}\right)+\mathcal O(\phi^3)
 \end{align}
The use of RNC is the reduction to parametrization independent magnitudes, i.e. Riemann tensor and its derivatives with the Christoffel symbols absent in our frame. In an analogy with general relativity, this is the free-falling frame where tidal effects reveal the geometry of the space-time manifold. 
In practice, there are no 3-point amplitudes~\footnote{They are reinstated however once we account for massive states.} and the interacting Lagrangian for 4-point reads:
\begin{align}\nonumber
\mathcal L_4^{\rm RNC}=&\frac16 R_{ha hb}\left(2h\partial h \varphi^a\partial\varphi^b-(\partial h)^2\varphi^a\varphi^b-h^2\partial\varphi^a\partial \varphi^b\right)\\
&+\frac16 R_{a bc d}\partial \varphi^a \varphi^b\varphi^c\partial\varphi^d
\end{align}
The first line gives the Feynman rule
\begin{align}\nonumber&
	\raisebox{-1cm}{\begin{tikzpicture}
		\draw [decorate,decoration=snake]  (-1/1.414,1/1.414) node[anchor=south east]{$\varphi^a(p_1)$}-- (0,0) ;
		\draw [decorate,decoration=snake] (-1/1.414,-1/1.414) node[anchor=north east]{$\varphi^b(p_2)$} -- (0,0);
		\draw [dashed]  (0,0)-- (1/1.414,1/1.414) node[anchor=south west]{$h(p_3)$};
		\draw [dashed]  (0,0)-- (1/1.414,-1/1.414) node[anchor=north west]{$h(p_4)$};
		\end{tikzpicture}}& \frac{iR_{ahbh}}3&\left(\begin{array}{c}
		(p_1+p_2)(p_3+p_4)\\+2p_1p_2+2p_3p_4
	\end{array}\right)
\end{align}
which evaluated on-shell is the sole diagram needed to compute $A_{WW\to hh}$ in this frame.
For 5-point vertexes, we have
\begin{align}\nonumber
\mathcal L_5^{\rm RNC}=&\frac{1}{12} (\nabla_h R_{\partial\varphi h h  \partial\varphi } +\nabla_h R_{\partial h \varphi\varphi \partial h}+2\nabla_h R_{\partial h \varphi h \partial\varphi}) 
\\
&+\frac{1}{12}( \nabla_h R_{\partial\varphi \varphi\varphi \partial\varphi}+2\nabla_\varphi R_{\partial\varphi h\varphi \partial\varphi}
)
\end{align}
 where the term $\nabla_{\varphi} R_{dh \varphi \varphi d\varphi}$  cancels due to the Riemann tensor asymmetry;
and with abuse of notation $V_{\varphi}=V_a \varphi^a$, similarly for $h$. For the 5-point amplitude, again due to the absence of 3-point vertexes, evaluating the Feynman rule that follows from the 5-point action yields the result (i.e. in this frame there is only the last diagram in fig.~\ref{fig5pt} to compute). 
Amplitudes for six or more particles in total do require a sum over diagrams and contain, in addition, poles which nevertheless can be derived from lower-point amplitudes, see~\cite{Cohen:2021ucp}.

A general EFT has also modifications in the pure gauge and fermionic sectors as e.g. a metric for the gauge kinetic terms \cite{Helset:2020yio,Hays:2020scx}; these are sub-leading in the high energy limit and for the observables here considered, although in practice they should be included in a complete analysis which we leave for future work.

\subsection{Experimental and theory constraints on curvature}\label{SecIIB}

Unitarity constrains the magnitude of curvature, and its derivatives, for a given c.m. energy $s$, to the 4-point level. Symbolically
\begin{align}\nonumber
	&2  {\rm\, Im}\left(\raisebox{-8mm}{\begin{tikzpicture}
		\draw [decorate,decoration=snake] (-1.732/2,1.732/2) -- (0,0);
		\draw [decorate,decoration=snake] (-1.732/2,-1.732/2) -- (0,0);
		\draw [decorate,decoration=snake] (0,0) -- (1.732/2,1.732/2);
		\draw [decorate,decoration=snake] (0,0) -- (1.732/2,-1.732/2);
				\filldraw (0,0) circle (2.5pt);
		\end{tikzpicture}}\right)
	+\raisebox{-10mm}{
		\begin{tikzpicture}
		\draw [decorate,decoration=snake] (-1.732/2,1.732/2) -- (0,0);
		\draw [decorate,decoration=snake] (-1.732/2,-1.732/2) -- (0,0);
		\filldraw (-0.1,0) circle (2.5pt);
	\draw [decorate,decoration=snake] ($(90:1 and 1)+(1,0)$) arc  (90:270:1 and 1);
	\draw (1.25,0) node {$\int\times d\Pi_{\rm LIPS}$};
	\draw [decorate,decoration=snake] ($(-90:1 and 1)+(1.5,0)$) arc  (-90:90:1 and 1);
			\filldraw (2.6,0) circle (2.5pt);
		\draw [decorate,decoration=snake] (2.5,0) -- (2.5+1.732/2,1.732/2);
		\draw [decorate,decoration=snake] (2.5,0) -- (2.5+1.732/2,-1.732/2);
		\end{tikzpicture}}\\ \nonumber
	&+\raisebox{-10mm}{
		\begin{tikzpicture}
		\draw [decorate,decoration=snake] (-1.732/2,1.732/2) -- (0,0);
		\draw [decorate,decoration=snake] (-1.732/2,-1.732/2) -- (0,0);
				\filldraw (0,0) circle (2.5pt);
		\draw [dashed] ($(90:1 and 1)+(1,0)$) arc  (90:270:1 and 1);
	\draw (1.25,0) node {$\int\times d\Pi_{\rm LIPS}$};
	\draw [dashed] ($(-90:1 and 1)+(1.5,0)$) arc  (-90:90:1 and 1);
			\filldraw (2.5,0) circle (2.5pt);
		\draw [decorate,decoration=snake] (2.5,0) -- (2.5+1.732/2,1.732/2);
		\draw [decorate,decoration=snake] (2.5,0) -- (2.5+1.732/2,-1.732/2);
		\end{tikzpicture}}+\cdots =0
\end{align}
 where the first partial wave for $W^+W^-$ gives
 \begin{align} \label{PartWaveUnit}
\left(\frac{ R_\varphi s}{16\pi}\right)^2+\frac12\left(\frac{R_h s}{8\pi}\right)^2\leq 1
\end{align}
where we have accounted for the amplitude being real. 
One can also select the $W^+ W^+$ channel, but the emphasis in here is on bounds which are sensitive to both curvatures simultaneously which helps to better close some corners in the curvature plane.

One can use these constraints to determine the theory cut-off in terms of curvature; however, here we turn this around to note that given that we have explored energies up to $s\sim v^2$ and no new states have showed up, we can set an upper limit on curvature.

This limit is super-seeded by experimental bounds from LHC which bound Higgs couplings. In the conventional parametrization, one has:
\begin{align}\label{FCtn}
F(h)^2=1+2a\frac{h}{v}+b\frac{h^2}{v^2}+\mathcal O(h^3)
\end{align}
which gives a curvature around the origin
\begin{align}\label{Rinab}
	v^2\left(R_\varphi(0),R_h(0)\right)=\left(1-a^2, -(b-a^2)\right)
\end{align}
itself related to amplitudes as, substituting~(\ref{Rdef},\ref{EqRabcd},\ref{EqRahbh}) on~(\ref{AWW},\ref{AWh}),
\begin{align} 
			\mathcal A_{W^+_1W^+_2\to WW}=&s_{12}R_\varphi \label{ExplWW}\\
			\label{ExplWh}
			\mathcal A_{W_1^+W_2^-\to hh}=& - s_{12}R_h
\end{align}

\begin{figure}[h]
	\includegraphics[width=\linewidth]{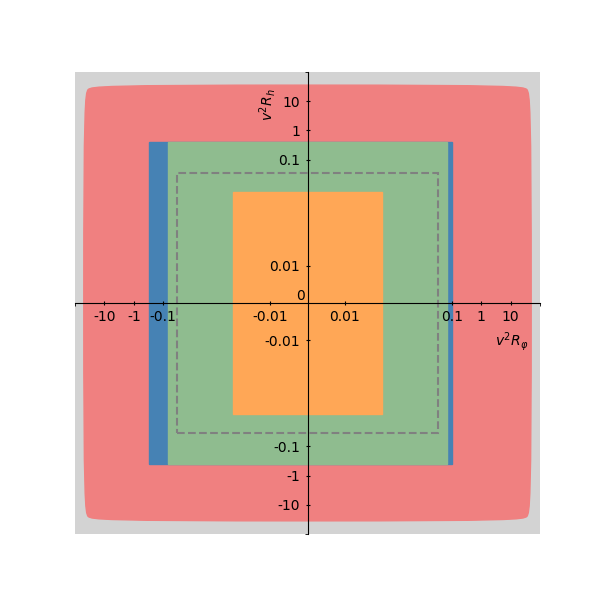}
	\caption{Theoretically (grey), and experimentally (up to blue) excluded (up to 95\% confidence level) regions of the curvatures $R_h, R_{\varphi}$ which are related to electroweak amplitudes as in eqs (\ref{ExplWh},\ref{ExplWW}); and sensitivity limits of future colliders (HL-LHC, up to green; FCC, up to orange), also up to 95\% confidence level. See text for detail. The plot scales linearly within the dashed box, and logarithmically outside.
	\label{Fig:THEXPBds}}
\end{figure}

Translating bounds on the coefficients from present and future measurements into curvature\footnote{Implicit in the relation of curvature and amplitudes is the assumption of no pure gauge sector modification justifiable given stringent constraints from LEP.}, we present the plot in fig.~\ref{Fig:THEXPBds}. The value in both sets of constraints is to put into context how much of the theory-consistent curvature space have we explored experimentally. 

From the outer-most to inner-most region of fig.~\ref{Fig:THEXPBds}: the (outer-most) grey region is excluded due to unitarity; up to the blue region is excluded by current LHC bounds (the region is translated from bounds on $a$ in \cite{ATLAS:2019nkf}, and $b$ in \cite{CMS:2021yqw}); finally, up to the green and orange (inner-most) regions we present expected exclusion limits for HL-LHC and FCC respectively. The projected bounds on $R_{\varphi}, R_h$ are derived using sensitivity predictions of $a$ (HL-LHC, \cite{Cepeda:2019klc}; FCC-ee, \cite{FCC:2018evy}); and $b$ (\cite{Bishara:2016kjn} for both HL-LHC and FCC-hh), around their SM values. All uncertainties and projected sensitivities are displayed at the $95\%$ confidence level; where multiple sensitivity estimates are given, the most conservative is selected. Note that HL-LHC bounds used here predate the LHC ones so that the seemingly marginal improvement is likely  an underestimation.

\section{Correlation of curvature in SMEFT}\label{SecIII}

In the linear realization and to first order (with our assumption of $O(4)$ invariance) we have:
\begin{align}
	R_\varphi=R_h
\end{align}
Which is to say the coefficients of $s$ in the 4-point amplitudes for $W^+W^+$ scattering and $W^+W^-\to hh$ in eqs~(\ref{ExplWh},\ref{ExplWW}) are anti-correlated. Correlations do appear in the linear parametrization of SMEFT in HEFT~\cite{Brivio:2013pma} in line with what we find here, nonetheless in this section we go into some length of how this can be derived to display the utility of a geometric language.

A simple argument to show there is a correlation, if a bit more abstract, is to use Riemann normal coordinates and custodial symmetry around the $O(4)$-symmetric point - which admits Cartesian coordinates. In this frame, the metric reads
\begin{align}
	G_{ij}(\phi)= \delta_{ij} + \frac13R_{iklj}\phi^k\phi^l+\mathcal O(\phi^3)
\end{align}
 and a linear realization of $O(4)$ symmetry dictates that the Riemann tensor be of the form
$R(\delta_{il}\delta_{kj}-\delta_{kl}\delta_{ij}) $, with a single unknown $R$. A transformation from Cartesian to polar coordinates then reveals $R_h=R_\varphi$.

The collapse of the two curvatures into a single one can also be derived matching the two EFTs:
\begin{align}\nonumber
	\frac{\left(\partial h^2+F^2\partial\varphi^2\right)}{2}=&K\left(\frac{H^\dagger H}{M^2}\right) (\partial H^\dagger H)^2\\&+G\left(\frac{H^\dagger H}{M^2}\right) D_\mu H^\dagger D^\mu H
\end{align}
where it should be understood from a general SMEFT action, we transformed to a basis where the Higgs singlet is canonically normalized.

This exercise yields, to order $M^{-4}$
\begin{align}
R_\varphi&=-3\frac{G'(0)}{M^2}+\frac{H^\dagger H}{M^4} \left(2(G'(0))^2-\frac 52 G''(0) \right) 
\\
R_h&=-3\frac{G'(0)}{M^2}+\frac{H^\dagger H}{M^4} \left(4(G'(0))^2-5 G''(0) \right) 
\end{align}
which also reveals the correlation is lost at order $M^{-4}$.

Finally, and in a direct connection with observables, one can compute the amplitude which has been used to define our curvature, the computation  itself getting rid of any field redundancy. Take the non-canonically normalized action 
\begin{align}\label{SMEFTAmp}
	\mathcal L=\frac12 \frac{c_{H\Box}}{M^2} (\partial_\mu H^\dagger H)^2+\frac{c_{HDD}}{M^2} H^\dagger H D_\mu H^\dagger D^\mu H
\end{align} 

\begin{figure}
\includegraphics[width=\linewidth]{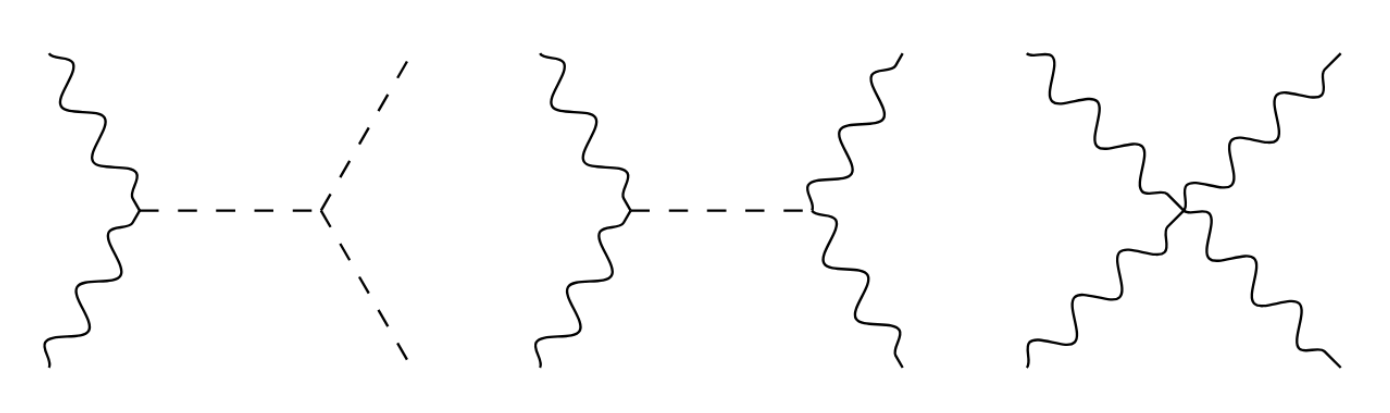}
\caption{\label{figSMEFTAmp} A selection of diagrams for the $WWhh$ and $WWWW$ amplitudes with the action in eq.~(\ref{SMEFTAmp})}
\end{figure}

After normalization of the theory, computation of diagrams such as those shown in fig.~\ref{figSMEFTAmp}, where we note that in this frame there is a $h^3$ coupling that scales with $s$ and must be accounted for, yields
\begin{align}
	\Amp{W^+W^+\to W^+W^+}=\frac{s}{M^2}\left(c_{H\Box} - c_{HDD}\right)\\ 
	\Amp{W^+W^-\to hh}=-\frac{s}{M^2}\left(c_{H\Box} - c_{HDD}\right)
\end{align}
and hence the direct connection with SMEFT geometry as 
\begin{align}\label{SMEFTR}
\left(R_\varphi\,,\,R_h\right)=\frac{1}{M^2}\left(c_{H\Box}-c_{HDD} , c_{H\Box}-c_{HDD} \right)\,.
\end{align}

\section{Models as probes into HEFT}\label{SecIV}
Recent study of EFT has shown that UV completion might impose extra constraints on an otherwise seemingly valid EFT, as is the case of positivity constraints~\cite{Adams:2006sv}. It should be said that these constraints on the curvatures themselves $R_h$ and $R_\varphi$ do not restrict their sign, but reveal the need for doubly-charged states if the curvature is negative~\cite{Falkowski:2012vh}. It is for these reasons that this section looks at models and introduces two new representations under $O(4)$ as
\begin{align}
	{\bf h}&: \quad 4\quad  {\rm of }\quad O(4) \\
	 \Phi&: \quad 9 \quad {\rm of } \quad O(4) \,\, ({\rm traceless\,\,symmetric})\\
	 S&: \quad 1\quad {\rm of } \quad O(4)
\end{align}
with the results of positivity constraints suggesting $S$ and $\Phi$ will produce positive and negative curvature respectively.
Note that $\bh$ is the Higgs doublet $H$ in a real representation as 
\begin{align} \label{Hbfhconnect}
	\left(\tilde H ,H\right) &= \hat \sigma_I  \frac{{\bf h}^I}{\sqrt2}
\end{align}
with $\tilde H=\epsilon H^*$ and $\sigma^I=(\sigma^i,1)$ with $\sigma^i$ the Pauli matrices.
We consider the addition of a 9 and a 1 separately with respective actions
\begin{align}
	\mathcal L_S=\frac12 D_\mu \bh^TD_\mu\bh+\frac12 (\partial S)^2-V(\bh,S^2)
\end{align}
\begin{align}
	\mathcal L_{\Phi}=\frac12 D_\mu \bh^TD_\mu\bh+\frac12 {\rm Tr}\left(D_\mu \Phi D^\mu\Phi\right)-V(\bh,\Phi)
\end{align}
The key distinction is whether $\vof{\Phi}=0$ or not, which depends on the sign of its mass term and its mixing as induced by the potential.

\subsection{ Only $h$ acquires a vev, SMEFT case}

In this subsection we momentarily restrict the $O(4)$ symmetry to $SO(4)$ to allow for tri-linear couplings.
First for the singlet $S$ case, we take a potential as 
\begin{align}
	V=-\frac{g_*m_S}{2}  S \,\bh^2+\frac{m_S^2}{2}S^2+\frac{m_\bh^2}{2}\bh^2
\end{align}
extra terms allowed by the symmetry will give controlled corrections to the result and we neglect them. Integrating the field $S$ at tree level returns
\begin{align}
	\mathcal L_{\rm eff}&=\frac{1}{2} \frac{g_*m_S}{2}\bh^2\frac1{\partial^2+m_S^2}\frac{g_*m_S}{2}\bh^2\\ &=\frac{g_*^2}{2}(H^\dagger H)^2+\frac{g_*^2}{2m_S^2}(\partial (H^\dagger H))^2+\mathcal O \left(\partial^4\right)
\end{align}
then via eq.~(\ref{SMEFTR}) 
\begin{align}
	\left(R_{\varphi},R_h\right)=\left(\frac{g_*^2}{m_S^2},\frac{g_*^2}{m_S^2}\right)
\end{align}
i.e. positive curvature for the singlet case, as expected.

Along the same lines, the potential for the symmetric representation is
\begin{align}
V=-\frac{g_*m_\Phi}{2} \bh^T \Phi \bh+\frac{m_\Phi^2}{2}\Phi^2+\frac{m_\bh^2}{2}\bh^2
\end{align}
The integration now returns, to dimension six:
\begin{align}
\mathcal L_{\rm eff}=&\frac{g_*^2}8 {\rm Tr}\left[\left(\bh \bh^T-\frac{\bh^2}{4}\right) \frac{m_\Phi^2}{\Box +m_\Phi^2}\left(\bh \bh^T-\frac{\bh^2}{4}\right)\right]\\ \nonumber
=&\frac{3g_*^2}{8}(H^\dagger H)^2+\frac{g_*^2}{m^2_\Phi}\left(H^\dagger H DH^\dagger DH+\frac{(\partial H^\dagger H)^2} {8}\right)
\end{align}
where $\Box=D_\mu D^\mu$ and one has that the operator does yield negative curvature:
\begin{align}
\left(R_{\varphi},R_h\right)=\left(-\frac{3g_*^2}{4m_\Phi^2},-\frac{3g_*^2}{4m_\Phi^2}\right)\,.
\end{align}

\subsection{Both $\Phi$ and h break the symmetry, HEFT/SMEFT quotient space}

As we will show, this case does not belong in SMEFT and stands as a representative of quotient space. We take the extension of a mexican hat potential for two fields as:
\begin{align}\nonumber
V(\Phi)=&-\frac{\vec{m}^2}{2}\cdot\left(\begin{array}{c}
\bh^2\\\Phi^2
\end{array}\right)+\left(\begin{array}{c}
\bh^2\\\Phi^2
\end{array}\right)^T\frac{\lambda}{8}\,\left(\begin{array}{c}
\bh^2\\\Phi^2
\end{array}\right)\\
&-\frac{\tilde \lambda}{8} \bh^T\Phi\Phi\,\bh +\frac{\tilde\lambda_{\Phi}}{8} {\rm Tr}\left(\Phi \Phi \Phi \Phi\right)
\end{align}
with ${\vec m}^2$ a 2-vector and $\lambda$ a $2\times2$ symmetric matrix. 
Since $\Phi$ acquires a vev, we take $\tilde \lambda>0$ which triggers $O(4)\to O(3)$ and preserves custodial symmetry. Linear terms in the fields are absent, contrary to the previous case which since we restore $O(4)$ in place of $SO(4)$. The key question as will be shown is to consistently compute particle couplings and masses from an explicit potential.

The Goldstone boson Lagrangian and couplings to the radial singlet modes $\delta h$, $\delta \Phi$ read:
\begin{align}
	\mathcal L
	=&\frac12\left((v_\bh+\delta h)^2+C_9(v_\Phi+\delta\Phi)^2\right)\frac{g_{ab}}{v^2}D^\mu\varphi^a D_\mu\varphi^b
\end{align}
where 
\begin{align}
	C_9&=\frac{2\times 4}{4-1}\,,&
	v^2&=v_{\bh}^2+C_9v_\Phi^2,& 
	\sin\beta&=\sqrt{C_9} \frac{v_\Phi}{v},
\end{align} and
\begin{align}
	\vof{\bh}&=\left(\begin{array}{c}
	0\\0\\0\\v_\bh
	\end{array}\right) &
	\vof{\Phi}&=\frac{v_\Phi}{2\sqrt{3}}\left(\begin{array}{cccc}
	1&&&\\
	&1&&\\
	&&1&\\
	&&&-3
	\end{array}\right)
\end{align}
the generalization of $C_9$ to $SO(N)$ being $C_{N(N+1)/2-1}=2N/(N-1)$.
Take the mixing for the singlet radial modes $\delta \bh$ and $\delta \Phi$ as (note that no other field in $\Phi$ or $\bh$ is a singlet of $SO(3)$ so we know these two only mix among each other):
\begin{align}
\left(\begin{array}{c}
\delta\bh\\
\delta \Phi
\end{array}\right)=\left(\begin{array}{cc}
\cos\omega & -\sin\omega\\
\sin\omega &\cos\omega
\end{array}\right)\left(\begin{array}{c} h\\\tilde h
\end{array}\right)
\end{align}
Putting the above back in the Lagrangian for the Goldstones and taking $h$ to be the lightest singlet, one obtains in our basis of eq.~(\ref{OurL},\ref{FCtn}) 
\begin{align}
a&=c_\omega c_\beta+\sqrt{C_9}s_\beta s_\omega&
b=&c_\omega^2+C_9 s_\omega^2
\end{align}

Note that the limit of no mixing gives $b=1$ and a parametrization of the curvature $R_h=-R_\varphi$ orthogonal to the SMEFT with a potential new road to the SM.
The question to be answered is then: can one take $\omega=\beta=0$ while
keeping $m_{\tilde h}\gg m_h$ and maintaining perturbativity?

To answer this question we should express $\omega$ and $\beta$ in terms of physical masses and couplings, then use eq.~(\ref{Rinab}) to substitute and find curvature as a function of physical masses and couplings. In practice we have to solve for the potential.
The value of the fields that minimize $V$ can be read off after rearranging as
\begin{align}
V(v_{\bf h},v_\Phi)= \left({\vec v}^2-2\hat\lambda^{-1}{\vec m}^2\right)^T\frac{\hat\lambda}{8}\left(\vec{v}^2-2\hat\lambda^{-1}{\vec m}^2\right)
\end{align}
with
\begin{align}
	{\vec v}^2&=2\hat\lambda^{-1}{\vec m}^2 & 
	\hat\lambda&=\lambda+\left(\begin{array}{cc}
	&-3\tilde\lambda/8\\
	-3\tilde\lambda/8& 7\tilde \lambda_\Phi/12
	\end{array}\right)
\end{align}
Next, expanding around the vevs we find the mass matrix for the singlets $\delta \bh,\delta\Phi$ as
\begin{align} 
M^2=&{\rm Diag} (v)\,\hat \lambda\, {\rm Diag} (v)=U\,{\rm Diag}(m_h^2,m_{\tilde h}^2)\,U^T 
\end{align}
with ${\rm Diag} (v)=\delta^{ij}v_j$. The aim is to express $\omega,\beta$ as $\omega(m_h,m_{\tilde h}, \hat\lambda,v),\beta(m_h,m_{\tilde h}, \hat\lambda,v)$, which can be done by taking the determinant of the mass matrix
\begin{align}  
{\rm det}(M^2)&=v_\bh^2 v_\Phi^2 {\rm det}(\hat\lambda)=m_h^2 m_{\tilde h}^2 
\end{align}
and combining the eigenvector equations into
\begin{align}
	\sin(2\omega)=\frac{2v_\bh v_{\Phi}}{m_{h}^2-m_{\tilde h}^2}\hat\lambda_{\bh\Phi}
\end{align}
to obtain 
	\begin{align}
		\sin(2\omega)&=\frac{2m_hm_{\tilde h} }{m_{h}^2-m_{\tilde h}^2}\frac{\hat\lambda_{\bh\Phi}}{\sqrt{\det(\hat \lambda)}}\\
	\sin(2\beta) &=\sqrt{C_9}\frac{2m_hm_{\tilde h}}{v^2\sqrt{\det\hat\lambda}}
	\end{align}
	No obstacle prevents taking $\omega\to 0$ with $\hat\lambda_{\bh \Phi}\to 0$, but it is evident that $\beta$ cannot be arbitrarily close to zero while keeping $\tilde h$ massive and respecting unitarity. Qualitatively then, we have a minimum attainable curvature as:
	\begin{align}
	\left(v^2R_{\varphi}\geq \frac{3m_h^2m_{\tilde h}^2}{8\pi^2 v^4 }\,,\,\, v^2 R_{h}\leq  -\frac{3m_h^2m_{\tilde h}^2}{8\pi^2 v^4 }  \right)
	\end{align}
where we took the unitarity bound on $\hat \lambda$ that follows from the 4-pt amplitude for $\delta h$ and $\delta\Phi$, see e.g.~\cite{Lee:1977eg}. This result, being proportional to the extra state mass, yields a naive cut-off $R=\frac{4\pi}{\Lambda^2}$ with inverse dependence on the new physics scale:
\begin{align}
	\frac{\Lambda^2}{v^2} \sim \frac{(4\pi)^3}{\lambda_{\rm SM}} \frac{v^2}{m_{\tilde h}^2}
\end{align} 
	so that the largest cut-off, or the closest to the SM couplings one can get, is attained for the lowest new physics scale. How low this scale can be while still being able to assume an EFT applies can be estimated from the amplitude for $W$ scattering, mediated by the singlets in the full theory
	\begin{align}
		-\mathcal A =\frac{s}{v^2}\left(1-c_\beta^2\frac {s}{s-m_h^2}-s_\beta^2\frac{s}{s-m_{\tilde h}^2}\right)+(s\to t)
	\end{align}
	
\begin{figure}
	\includegraphics[width=\linewidth]{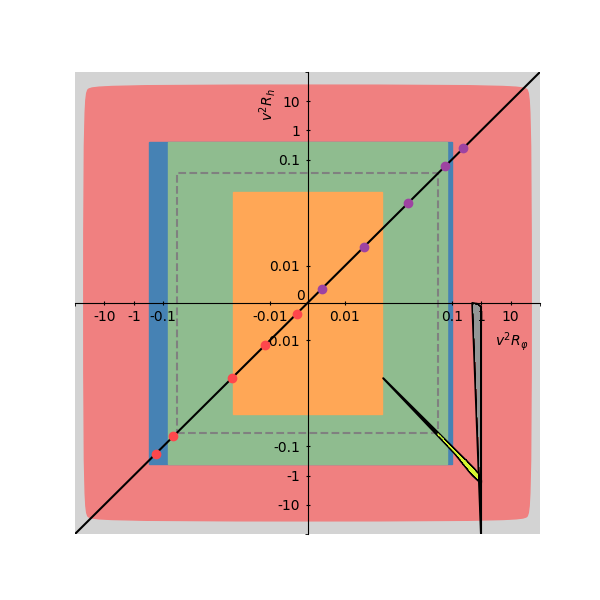}
	\caption{\label{FigSym} Range of curvature for SMEFT and quotient theories, on the same background as Fig. \ref{Fig:THEXPBds}. Two quotient theories are plotted: the yellow region shows curvature for the symmetric representation with $\vof{\Phi}\neq0$, and the dark-grey region shows a hyperbolic manifold (see sec. \ref{SecV}). The black line shows SMEFT curvature; on which the purple and red dots represent the singlet and the symmetric representation with $\vof{\Phi}=0$ examples from sec. \ref{SecIV} respectively. The outer-most to inner-most dots are evaluated with coupling $g_*=1$ and heavy singlet mass: 500 GeV, 1 TeV, 1.5 TeV, 2 TeV and 4 TeV.}
\end{figure}

	The plot in fig.~\ref{FigSym} shows the region in the curvature plane that the models discussed in this section cover. In particular for the minimum mass of the extra singlet we take the limit of $m_{\tilde h}\gtrsim 350$ GeV from \cite{ATLAS:2017uhp} as reference.

\section{Manifolds}\label{SecV}
The above HEFT cases fall into the category of manifolds with a singularity, as one can see by integrating out heavy states~\cite{Cohen:2020xca}. In contrast, one can also have that no $O(4)$-symmetric point is present and the manifold is smooth at every point. This section visualizes both types of manifolds, together with those that admit a SMEFT description. Consider (higher dimensional) cylindrical coordinates, the gauge symmetry acts rotating along the axis and orthogonal to this rotation we have a cylindrical radial coordinate $\rho$ and a `height' $z$. Our manifolds are hypersurfaces within this 5d space parametrized by $h$ and $\varphi^a$ 
\begin{align}
	(\rho(h)u(\varphi),z(h))
\end{align}	
With a line element:
\begin{align}
	d\ell^2=\left(\left(\frac{d\rho}{dh}\right)^2\pm\left(\frac{dz}{dh}\right)^2\right)dh^2+\rho(h)^2 du^2
\end{align}
which defines the 4-d metric,
where the plus sign is for Euclidean 5d space and the minus for the metric with a $(-1,1,1,1,1)$ signature.
In our basis, eq.~(\ref{OurL}), $dh^2$ has unit coefficient which can always be attained by a field redefinition. In terms of geometry, the singlet Higgs field $h$ equals distance in field space for fixed $u$. From the equation above and our basis it also follows that $F(h)=\rho(h)/v$ with $F(0)=1$ giving $\rho(0)=v$. For convenience let us define $\theta = (h+h_0)/f$ with $f$ a new physics scale.

The most symmetric manifolds are $S^4$, $R^4$ \& $\mathcal H^4$ which are parametrized in our basis as
\begin{align}
	S^4& &&(f\sin(\theta)u , f\cos(\theta))\\
	R^4& &&((h+v) u , 0)\\
\mathcal	H^4& &&(f\sinh(\theta)u , f\cosh(\theta))
\end{align}
and yield constant (field-independent) curvature:
\begin{align}\nonumber
& && R_\varphi,&&R_h\\
S^4,\mathcal H^4&&& 
\pm\frac{1}{f^2},&&\pm\frac{1}{f^2}
\end{align}
 while the $f\to \infty$ limit yields $R^4$ which corresponds to the SM. Indeed these manifolds can be described in SMEFT and correspond to Composite Higgs Models~\cite{Panico:2015jxa} or negative curvature models~\cite{Alonso:2016btr}. 
 
 \subsection{quotient space theories with a singularity}
A one-parameter deformation of the manifolds above takes us into quotient space with a singularity at the origin:
\begin{align}
	{\rm deformed\,} S^4& &&\left( fs_{\gamma \theta}u, \int dh \sqrt{1-\gamma^2c^2_{\gamma \theta}} \right)\\
	{\rm deformed\,} \mathcal H^4& &&\left(  f sh_{\gamma \theta}u,\int dh \sqrt{\gamma^2ch^2_{\gamma \theta}-1}\right)
\end{align}
where $s_{\gamma\theta}=\sin(\gamma \theta)$ and
 the singularity is made evident by the curvature
\begin{align}\nonumber
& && R_\varphi &&R_h\\\label{DfS4}
{\rm deformed\,} S^4&&& \frac{1-\gamma^2}{f^2s^2_{\gamma\theta}}+\frac{\gamma^2}{f^2},&&\frac{\gamma^2}{f^2}\\
{\rm deformed\,} \mathcal H^4&&& \frac{1-\gamma^2}{f^2sh^2_{\gamma\theta}}-\frac{\gamma^2}{f^2},&&-\frac{\gamma^2}{f^2}\label{DfH4}
\end{align}
since the origin, and would-be-$O(4)$ invariant point, $\theta=0$, returns $R_\varphi=\infty$. This singularity is present for any $\gamma\neq \pm 1$ which seemingly presents a way to approximate the SM by sending first $f\to\infty$ while keeping $f s_{\gamma\theta_0}(f sh_{\gamma\theta_0})=v$ constant, then $\gamma\to 1$. Indeed in this limit, $\partial^n R\propto (1-\gamma^2)$ and contributions to amplitudes of an arbitrary number of particles cancel. Nonetheless and quite relevantly in this limit, the singularity is just a field distance $v/\gamma$ from the vacuum $h=0$.
 The model in the section above with a symmetric representation taking a vev also belongs to the quotient theories with singularities, yet it showed that the SM point cannot be reached. So it could be that the deformed manifolds have no UV completion, yet from low energy we see no indication for it. This highlights the need for a  bound based purely in the EFT perspective to comprise all possibilities. 

\begin{figure}[ht]
\centering
\includegraphics[width=.45\textwidth]{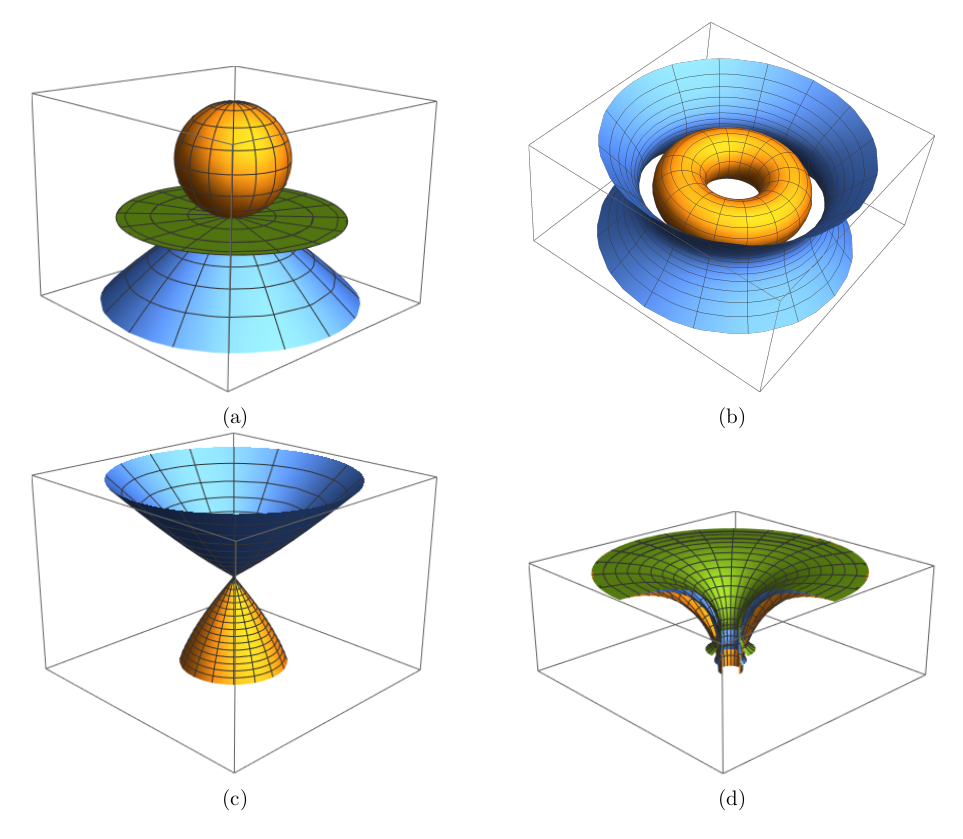}
\caption{\label{figAll} Examples of manifolds which belong in SMEFT (a), or in quotient space (b,c,d) with the gauge symmetry action being rotation around the $z$ axis. SMEFT manifolds in (a) correspond to: Composite models (yellow), the SM (green), and negative curvature models (blue). quotient manifolds (b,d) are smooth, while (c) present a singularity and both (c,d) are in a class which resembles the SM around the vacuum. For (d), part of the manifolds has been cut out for better visualization.}
\end{figure}

\subsection{Smooth quotient theories}

On the other hand, one could have smooth manifolds in quotient space, $\rho\neq 0\, \forall \,h$; we take here as examples a torus and a hyperbola (in Euclidean space)
\begin{align}
	{\rm torus}& & &(  (\rho_0+fc_{\theta})u, fs_\theta)\\
	{\rm hyperbola}& & &( (\rho_0+fch_{\hat \theta})u, f sh_{\hat \theta})
\end{align}
where $\hat \theta=(\hat h(h) +\hat h_0)/f$ with $(dh/d\hat h)^2=sh_{\hat\theta}^2+ch_{\hat\theta}^2$ as follows from our normalization in Euclidean 5d.
In terms of curvature, these manifolds give:
\begin{align}\nonumber & && R_\varphi && R_h	\\
{\rm Torus} & && \frac{\cos(\theta)^2}{v^2},& &\frac{\cos(\theta)}{fv}\\
{\rm Hyperbola} & &&  \frac{ch_{\hat\theta}^2}{(ch_{\hat\theta}^2+sh_{\hat\theta}^2)v^2},& &\frac{-ch_{\hat\theta}}{(ch_{\hat\theta}^2+sh_{\hat\theta}^2)^2fv}
\end{align}
We see that the hyperbola does not go through the zero curvature point for any value of $f, \theta$, always keeping a distance as the explicit model in the previous section did. The torus however for $\theta=\pi/2$ does have both curvatures vanish, yet by construction the manifold is not $R^4$. Visually, for this point we are sitting atop of the torus and for its first two derivatives it does resemble a plane, but its third derivative is non-vanishing and indeed $ R'_h=1/f^2v$ which is bounded from below given $\rho_0>f$ and for $\theta=\pi/2$, $v=\rho_0$.

This nonetheless illustrates the possibility of manifolds that do look locally like the SM to the $n$th derivative, yet do not go through the origin. Let us take on such set of manifolds labelled by $n$ 
\begin{align}\label{cpxp}
	F_{(n)}(h)&=1+\frac{h}{v}+c_n\left(\frac{h}{v}\right)^n & |c_n|&>\frac{(n-1)^{n-1}}{n^n}
\end{align} 
The manifolds associated with these $F_n$ for $n=3,4,5$ are plotted in fig.~\ref{figAll} and they resemble a plane and hence the SM ever more accurately for increasing $n$ around $h=0$. 

It should be underlined that we do not know an UV completion that would yield this type of EFT, as opposed to quotient theories with singularities.

\section{Obstacles in the road to the SM}\label{SecVI}

We have encountered HEFT/SMEFT quotient theories which either come from smooth manifolds with no $O(4)$-invariant point, or manifolds which get arbitrarily close to the would be $O(4)$-invariant point, but the point itself is singular.


A number of UV complete theories yield quotient theories with singularities at the origin. From working out an explicit example, we have seen that these can only get within a finite distance of the SM point. This explicit computation relied on knowledge of the full theory, but here we attempt to give an argument as to why quotient theories are not a road to the SM model in purely low energy grounds.

Let us turn to semi-classical arguments. Consider the Higgs field as sourced by a probe particle $i$ localized in a region $\sigma_x$ and with a mass $m_i> m_h$. This configuration is, of course, short lived yet for times smaller than the decay rates one might consider such system. The renormalizable linear realization gives an equation of motion~\footnote{The spin 0, 1 case has an extra $h/v$ times the source which we dropped},
\begin{align}\label{EoM}
	(-\Box-m^2_h)h(x)=\frac{m_i}{v} J_i(x)
\end{align}
where
\begin{align}
	&{\rm Spin\,1/2} &	J_i&= \langle i| \bar\psi \psi |i\rangle\\
	&{\rm Spin\,1} &J_i&=- \langle i|  m_i V_\mu V^\mu|i\rangle
	\end{align}
and the particle state is 
\begin{align}
	|i\rangle=& \int \frac {d^3 p}{(2\pi)^3} \Psi(p) \frac{a_{i,p}^\dagger}{\sqrt{2E_p}}|0\rangle
\end{align}
Away from the localized source the field is
\begin{align}
	h(r>\sigma_x)&=\frac{m_i}{v}\int \frac{d^4xd^4q}{(2\pi)^4}\frac{e^{iq(x-y)}\hat J_i(\vec x)}{q^2-m^2}\\
	&\simeq -\frac{m_i}{v}\frac{e^{-m_hr}}{4\pi r}
\end{align}
where in the second line we assumed that the current $J_i$ is the same as the probability density, as we shall see justified in the non-relativistic limit.

Consider now the candidate quotient theories that resemble the Standard Model to a high degree, examples given in the previous section are the functions $F_{(n)}$ as given in~(\ref{cpxp}) or the deformed $S^4, \mathcal H^4$ theories~(\ref{DfS4},\ref{DfH4}). The solution above should be a good first approximation certainly for large distances $r>1/m_h$ where the field value is exponentially close to the vacuum value. However, at shorter distances if our candidate theories truly present a limit in which the SM couplings are recovered, the solution should still be a good approximation. The field value nonetheless increases with decreasing distance and if there is a singularity, in this SM limit, it is just a distance $v/\gamma\simeq v$ away in field space. Conversely, for smooth quotient theories, even if our series example $F_{(n)}$ resembles the SM locally around the vacuum, the corrections in eq.~(\ref{EoM}) read $1+n c_n (h/v)^n$ with $nc_n\sim 1$ for $n\gg1$ and would dominate over the SM for $h\sim v$. This is indeed the same condition for both types of theories and yields a naive minimum distance or cut-off

\begin{align}\label{naive}
	&\frac{h(\sigma_x<r<m_h^{-1})}{v}\simeq \frac{m_i}{v}\frac{1}{4\pi v r} \\ &\frac{h(r_0)}{v}\sim1 \quad {\rm for}\quad \frac{1}{r_0}\equiv\Lambda\sim 4\pi v \frac{v}{m_i}
\end{align}

This points at a cut-off an inverse coupling factor higher than other estimates based on pertubative unitarity. 
Nevertheless, quantum mechanics has something to say about our implicit assumption $\sigma_x<r_0$. Indeed $r_0\sim (m^2_i/4\pi v^2) m_i^{-1}$ is smaller than the inverse mass of a particle for perturbative couplings (which is the case for the SM) but in order to localize the particle in a distance smaller than the inverse mass, the uncertainty principle dictates a range of momenta that extends to the relativistic regime. In this high energy limit, our current $J_i$ suffers a relativist factor $m/E$ suppression as explicit evaluation of the matrix elements shows when going beyond the non-relativistic approximation. For a fermion, one has
\begin{align}
	J_i(x)=\int\frac{d^3pd^3k}{(2\pi)^6}\frac{\bar u(k)u(p)}{\sqrt{2E_{p}2E_k}}e^{i(p-k)x} \Psi^*(k)\Psi(p)
\end{align}
 which implies that the space-integral over the source $J_i$ is suppressed and the field value at a distance $r>\sigma_x$ is
\begin{align}\label{Realhofr}
	\frac{h(\sigma_x<r)}v&= \frac{N(m_i\sigma_x)}{4\pi v r} \frac{m_i}{v^2}=\frac{N(\sigma_x m_i)}{r m_i} \alpha_i\\
	N(m_i,\sigma_x)&=\frac{\int d^3k (m_i/E_p) |\Psi(p)|^2}{\int  d^3k  |\Psi(p)|^2}\quad \alpha_i=\frac{m_i^2}{4\pi v^2}
\end{align}
which is the same result for spin $1/2$ and $1$.
This suppression implies that the pre-factor of $\alpha_i$ in the eq.~(\ref{Realhofr}) is at most order one, which would then require an order one $\alpha_i$ to probe $(h/v)\sim 1$. Note that this $\alpha_i$ will be at the edge of perturbative unitarity, although loop corrections will be supressed by $\sim1/(4\pi)$.

 As an estimate, we take a Gaussian distribution $\Psi\sim e^{-(p\sigma_x)^2/2}$ and evaluate the potential at a distance $r=2\sigma_x$ which encloses 95\% of the probability density to find that with $\alpha_i\sim 2$ the cut-off, or inverse distance, where we would probe $h\sim v$ would be $r_0=0.6m_i^{-1}$,
 \begin{align}
 \Lambda \sim \sqrt{\frac{8\pi\sigma_x m_i}{N(\sigma_x m_i)} }\Bigg|_{m_i\sigma\sim 0.3}v\simeq 2{\,\rm TeV} \,.
 \end{align}

The nature of EWSB and the question of whether a symmetric $O(4)$ point exits should be independent of the introduction of our probe particle $i$, although admittedly the fact that one would require couplings on the pertubative edge makes the above a rough estimate.

The naive scaling from eq.~(\ref{naive}) does, however, point towards the typical scale for non-perturbative effects. This is indeed the natural scale for answering non-local questions about our theory. While the detailed study of this effect will be presented elsewhere~\cite{InPrep} here we sketch the modifications in a well known non-perturbative effect, sphalerons, whose energy is
\begin{align}
	E_{\rm sph} \sim \frac{4\pi v}{g}
\end{align}
In particular, the topological argument by Manton~\cite{Manton:1983nd} has to do with a loop (parametrized by $\mu$) of mappings from the sphere  at spatial infinity to the vacuum manifold, characterized by our unit vector $u$, i.e. $u(\theta,\phi;\mu)$ and holds regardless of the Higgs singlet role. Nonetheless the boundary conditions {\it to find the energy} of the potential barrier have to be drastically changed in quotient theories. Indeed, the proposed field at the top of the barrier $\mu=\pi/2$ in~\cite{Manton:1983nd} is $(\bh =h(r) u)$
\begin{align}
	\bf h &= h(r)\left(\begin{array}{c}
		s_\mu s_\theta c_\phi\\ s_\mu s_\theta s_\phi \\
		s_\mu c_\mu (c_\theta-1) \\ s_\mu^2 c_\theta+c_\mu^2
	\end{array}\right)&
 {\rm B.C.}&\left\{
 \begin{array}{c}
 	h(0)=0\,,\\
 	h(\infty)=v\,.
 \end{array} \right.
\end{align} 
In particular, the condition at the origin, that the Higgs field go to its symmetry preserving $O(4)$ symmetric point, is demanded to remove dependence on angular variables of the Higgs doublet at the origin where $\theta,\phi$ are ill-defined. For quotient theories, it is clear that this does not apply given that an $O(4)$ point is absent or singular. One can introduce a radial dependent function on $u$ itself such that
\begin{align}
u(\theta,\phi,r\to \infty)&= u_\infty &
u(\theta,\phi,r\to 0)&\to u_0	
\end{align}
The boundary conditions on $h$ would naively be $h'(0)=0$. In either case, the quotient theory effect is an order one modification which serves as a handle to tell quotient theories apart from the Standard Model.

\section{Summary}

This work studied the quotient space HEFT/SMEFT, and the potential limits to recover the SM other than via SMEFT with the use of a geometric formulation.
 Explicit examples, which include perturbative UV complete models, can and will be told apart from the SMEFT case by future experiments via projection of measurements on the curvature plane defined from the $WW$ scattering and $WW\to hh$ amplitudes (see fig.~\ref{FigSym}). These examples of quotient space HEFT/SMEFT theories do not offer a limit to recover the SM and possess a finite cut-off. In contrast to these, quotient theories were formulated in sec.~\ref{SecV} which resemble the SM amplitudes for arbitrary precision and number of particles. While these theories look like the SM model around the vacuum, at a Higgs-singlet-distance of $\sim v$ they reveal their quotient space nature.
 Making use of semi-classical arguments to displace the Higgs field by $\sim v$, we find an argument for general theories in quotient space to be distinguishable from the SM when probing the theory at an energy (inverse distance) of at most $4\pi v/g_{\rm SM}$. Our discussion applies to quotient theories both with and without singularities (non-analyticities). The most pressing outstanding question is the characterization of experimental signatures that follow from the semi-classical arguments given here.

\section{Acknowledgements}

R. A. and M.W. are supported by the STFC under Grant No. ST/P001246/1.



\bibliography{GAEWSBLib.bib}

\end{document}